\documentclass[iop,apj,twocolappendix,twocolumn]{emulateapj}
\pdfoutput=1
\usepackage{amssymb}
\usepackage{amsmath}
\usepackage{natbib}
\usepackage{subfigure}
\usepackage{booktabs}
\newcommand{\bld}[1]{\mbox{\boldmath$#1$\unboldmath}}
\usepackage{color,ulem}

\begin{document}

\title{Theory and simulations of rotating convection}
\shorttitle{Rotating Convection}
\shortauthors{A.~J. Barker, A.~M. Dempsey \& Y. Lithwick}
\author{Adrian J. Barker$^{1,2}$, Adam M. Dempsey$^{1}$ and Yoram Lithwick$^{1}$}
\affil{$^1$
	  Center for Interdisciplinary Exploration and Research in Astrophysics (CIERA) \& \\ Dept. of Physics and Astronomy, Northwestern University, 2145 Sheridan Rd, Evanston, IL 60208, USA}
	  \affil{$^2$ Department of Applied Mathematics and Theoretical Physics, University of Cambridge, Centre for Mathematical Sciences, Wilberforce Road, Cambridge CB3 0WA, UK}
	  \email{ajb268@cam.ac.uk}

\begin{abstract}
We study thermal convection in a rotating fluid in order to better understand the properties of convection zones in rotating stars and planets.
We  first derive 
  mixing-length theory for rapidly-rotating convection, arriving at 
  the results of Stevenson (1979) via simple physical arguments.
The theory
  predicts  the properties of  convection as a function
of the imposed heat flux and rotation rate, independent of microscopic diffusivities.
 In particular, it predicts the mean temperature gradient; 
 the rms velocity and
temperature fluctuations; and the size of the eddies that dominate heat transport.
We test all of these predictions  with high resolution  three-dimensional hydrodynamical simulations
 of Boussinesq convection in a Cartesian box.
The results agree remarkably well with the theory across more than two orders of magnitude in rotation rate. 
 For example, the temperature gradient is predicted to scale as
the rotation rate to the 4/5th power at fixed flux, and the simulations yield $0.75\pm 0.06$.
We conclude that the mixing length theory is a solid foundation for understanding the properties of
convection zones in  rotating  stars and planets.
\end{abstract}
\maketitle

\label{firstpage}
\section{Introduction}
\label{sec:intro}

Rotating convection   occurs in the interiors of  many stars and planets. 
But there is no adequate theory  for it yet,
despite many decades of research. 
In order to determine the  structure of a non-rotating star or planet, one typically employs 
 mixing length theory 
 (e.g.~\citealt{BohmVitense1958}), which
 despite its crudeness, 
 accounts for the main structural feature of a convection zone:  a nearly uniform entropy.
 But it remains unclear whether mixing length theory applies quantitatively, and how it may be extended to treat more subtle effects such as rotation and overshooting.

Convection in astrophysical bodies---rotating or not---is  difficult theoretically  because the
flow is
 turbulent.  Compounding the difficulty,   microscopic 
 viscosities 
($\nu$) and thermal diffusivities ($\kappa$)
   are 
 typically extremely small in stars and planets, orders of magnitude smaller than  
 accessible by experiment or simulation \citep[e.g.,][]{Spiegel1971,Miesch2005}. 
  It is generally believed that  bulk properties
 of turbulence should be independent of microscopic diffusivities in the limit
 that these are extremely small.  But it remains unclear 
 if that belief is correct, and if it is,  whether
  experiments and simulations are adequately probing that limit.

  Rather than study the full problem in a star or planet, a common approach is to consider
 a simpler setup, Rayleigh-Benard convection (RBC), 
which has been extensively studied theoretically \citep[e.g.,][]{C1961,GrossmanLohse2000}, numerically, and experimentally (e.g.~\citealt{Ahlers2009}, and references therein). 
 In RBC, 
a  fluid layer is sandwiched
 between two horizontal plates. Convection is driven in the interior
 by holding the bottom plate  hotter  than the top.
 The fluid obeys the
 Boussinesq equations,  
  in which the density is constant, 
  the velocity field is incompressible, and temperature fluctuations give rise to vertical buoyancy
 forces \citep[e.g.][]{C1961}.  This is perhaps the simplest
 set of equations that self-consistently evolve turbulent convection.  

Because of its simplicity,  RBC is an ideal testbed for mixing length theory.  
The typical goal is to
 predict the heat flux through
the fluid given the temperature drop between the plates.  
But the mixing length prediction,  taken at face value, 
 wildly disagrees with experimental and numerical results.\footnote{In the absence of rotation, mixing length theory predicts 
 Nu $ = $ const$\times $Ra$^{1/2}$ \citep{Kraichnan1962}, where Ra (the Rayleigh number) quantifies the temperature
 drop and Nu (the Nusselt number) quantifies the  heat flux. 
  By contrast, experiments
 and simulations obtain an exponent 
 $\approx 1/3$ at high Ra, i.e., at small $\nu$ and $\kappa$
 \citep{Spiegel1971,ShraimanSiggia1990,Ahlers2009,2010NJPh...12k5002N}.
 The discrepancy increases as $\nu,\kappa\rightarrow 0$.
Note that the mixing length prediction  may 
be derived by dimensional analysis under the assumption that
 viscosity and thermal diffusivity play no
 role (\S \ref{mlt}).}
The reason for the discrepancy is that a naive application of mixing length theory
assumes  a constant temperature gradient throughout the fluid.  But
in reality,
most of the temperature drop 
occurs in extremely thin boundary layers near the top and bottom plates,
where fluid  velocities are nearly zero and heat is transported primarily by
conduction. 
In fact, 
the relationship between temperature 
drop and heat flux can be accounted for by considering only the behavior of
 the boundary layers \citep{Malkus1954}. 
For {\it rotating} RBC, the situation is  similar:  \cite{King2012}  
 extend the boundary layer analysis of \cite{Malkus1954} 
to include rotation, and thereby obtain
impressive agreement with experiments and simulations (see also \citealt{King2009,King2013a}). 

However, for the purpose of explaining convection in astrophysical bodies, 
it is the behavior of the bulk of the fluid that is primarily of interest---not the boundary 
layers. The boundary layers in RBC are very different from the boundary 
of a  convection zone in a star or planet. But one might expect that the turbulent dynamics
of the interior fluid will be similar in the two cases.  Therefore in this paper we focus on the
dynamics of the
 interior
fluid in an RBC-like system.  We shall show that not only do the properties of the interior fluid converge
in the limit of small diffusivities, but  they converge to the prediction
of the rotating mixing length theory  first proposed by \cite{Stevenson1979}.
 
The organization of this paper is as follows. We first set up the problem (\S \ref{Theory}), and then derive the predictions of
 mixing length theory (\S \ref{mlt}).  Next, we test the theory in detail with a suite of numerical
 simulations that are similar to---but slightly different from---standard rotating RBC  (\S \ref{Results}).  We also run some comparison simulations with standard
 RBC  (\S \ref{RBC}). We conclude with a summary and discussion (\S \ref{Conclusions}).
 
\section{Problem Setup}
\label{Theory}

We consider a convecting fluid  in a region of depth $H$ that carries thermal flux $F$ and rotates at rate $\Omega$.
Our goal is to determine 
  the    properties of the turbulently convecting fluid, 
 focusing in particular on the mean temperature gradient, the rms velocity
and temperature fluctuations, and the characteristic size  of the eddies that dominate the flux. 
We eschew the commonly used dimensionless numbers (Ra and Nu) because
  they depend on the temperature
drop across a  region that includes both convecting fluid and boundary layers, and
are thus a better diagnostic of the boundary layers than of the convective fluid.
(But see Table \ref{Table}, where we catalog these numbers in our simulations to 
facilitate comparison with prior work).

 Our setup is perhaps the simplest one possible:
 Boussinesq convection in which 
the rotation vector is aligned with gravity.
 But the physical argument  will be
  straightforward to extend to more complicated situations.
The Boussinesq equations are 
 \begin{eqnarray}
\label{eq1aa}
&& \nabla \cdot \boldsymbol{v} = 0,
\label{eq1b}
\\
\label{eq1a}
&&\left(\partial_{t} + \boldsymbol{v}\cdot \nabla\right)\boldsymbol{v}  +2\Omega
\boldsymbol{e}_{z}
\times\boldsymbol{v}= -\nabla p +  T\boldsymbol{e}_{z} ,
 \\
&& \left(\partial_{t} + \boldsymbol{v}\cdot \nabla \right)T = 0,
\label{eq1c}
\end{eqnarray}
(e.g.~\citealt{Spiegel1960})\footnote{
If one decomposes the temperature into a horizontally 
uniform part and a remainder, 
$T=T_0(z)+T_1(\bld{x},t)$, then
Eq. \ref{eq1c}
becomes
$ \left(\partial_{t} + \boldsymbol{v}\cdot \nabla \right)T_1 = -v_z{dT_0/ dz} $
while Eq. \ref{eq1a} remains unchanged, aside from replacing $T\rightarrow T_1$ (since
 $T_0$ is absorbed by $p$).}
which 
 express, respectively,  incompressibility, force balance, and heat balance.
Here, $\bld{v}$ is the velocity field;
$p$ is a pressure variable that  is determined
by incompressibility; 
 the gravitational
acceleration points downwards (in the $-\bld{e}_z$ direction);
the rotation vector (with magnitude $\Omega$) points upwards; 
 and $T$ is a scaled temperature.  
More precisely, $T=g\alpha (\delta T_{\rm true}/T_{\rm true})$, where $g$ is gravity, 
 $\alpha$ is a thermal expansion coefficient, and $\delta T_{\rm true}/T_{\rm true}$ is the fractional difference in the true temperature relative to a reference value.
 One may also  think of $T$ as the buoyancy variable, 
 since in the Boussinesq approximation  temperature only affects the equation of motion through buoyancy; as such,  $T$ has dimensions of acceleration. 

Equations \ref{eq1a} and \ref{eq1c} must be supplemented with  diffusive terms
(e.g.,  
 $\nu\nabla^2\bld{v}$ and $\kappa\nabla^2T$, where $\nu$ is viscosity and $\kappa$ is thermal
 diffusivity).
In most astrophysical applications, $\nu$ and $\kappa$ are extremely small.
 Thus, while they can affect dynamics on extremely small lengthscales, 
 that presumably has little effect on large-scale properties.
 Our results below will confirm this presumption.

\section{Mixing Length Theory of Rotating Convection}
 \label{mlt}
 
We assume (subject to verification below) that
the equations of motion without diffusivities are responsible for the large-scale dynamics
in the bulk of the convecting region.
We decompose the temperature  into a constant-gradient background and fluctuation
\begin{eqnarray}
T={\rm const}-N_*^2z+T_1 \ ,
\end{eqnarray}
where $N_*$ is a positive constant with dimensions of frequency. It is related
to the usual Brunt-Vaisala frequency $N$ via $N_*^2=-N^2$.
Linearizing
the Boussinesq equations  in $T_1$ and $\bld{v}$ yields the dispersion relation
for the growth rate $\gamma$:
\begin{equation}
\gamma^2= N_*^2{k_\perp^2\over k^2}-(2\Omega)^2{k_z^2\over k^2} \ ,
\label{eq:dr}
\end{equation}
where $k_\perp$ and $k_z$ are the components of the wavevector in the horizontal
and vertical directions.
Modes with $k_\perp> k_z(2\Omega/N_*)$ are exponentially unstable ($\gamma^2>0$), 
as they would be in a fluid without rotation. Although the Coriolis force stabilizes some modes, 
it cannot stabilize those whose phasefronts are sufficiently vertical.
But such modes obviously cannot continue to grow indefinitely.  Rather,  they 
 grow until nonlinearities cause  them to cascade. That is the basis for the mixing length
theory, described  in what follows. 

We focus here and in the remainder of this paper on the rapidly 
rotating limit, which we take to mean
\begin{equation}
2\Omega\gg N_* \ .
\end{equation}
The dispersion relation   implies that modes with $k_\perp/k_z> {\rm few \ }\times (2\Omega/N_*)\gg 1$ grow at the rate $\gamma\approx N_*$. 
Although all such modes are unstable, one expects that the modes
with the largest wavelengths are responsible for the bulk of the heat transport (or equivalently
that
   convection is dominated by the mode that transports the most heat -- e.g.~\citealt{Malkus1954,Stevenson1979}).
The largest wavelength in the vertical direction is the one that fits in a convection zone
of depth $H$:
\begin{equation}
k_z \sim {1\over H} \ .
\label{eq:kz}
\end{equation}
In the horizontal direction, the largest unstable mode then has
\begin{equation}
k_\perp \sim {1\over H}{2\Omega\over N_*} \ .
\label{eq:kperp}
\end{equation}
In  convective steady state the growth rate of such modes  equals their nonlinear
cascade rate, $\sim k_\perp v_\perp$,\footnote{
When employing the ``$\sim$'' symbol, we take  $v_\perp$, $v_z$, and $T_1$ as referring to R.M.S. values.} since that is the rate at which fluid is mixed across horizontal distance $1/k_\perp$.
We thus infer that
\begin{equation}
N_*\sim k_\perp v_\perp.
\label{eq:relation1}
\end{equation}
We wish to relate the temperature gradient to the thermal flux $F=\left<v_z T \right>$, where
angled brackets denote averaging over a horizontal plane; i.e.,
\begin{equation}
F \sim v_z T_1 \  ,
\end{equation}
since hot fluid typically rises and cold fluid sinks.
To complete the calculation requires an estimate for $T_1$.  As hot plumes rise, they carry the background temperature upwards for a time $1/\gamma\sim 1/N_*$ before cascading, 
i.e., 
$T_1\sim {d\left<  T\right>/ dz}\times {v_z/ N_*}$. 
 Since $d\left<  T\right>/ dz=-N_*^2$, we have
\begin{equation}
T_1
\sim v_zN_* \ .
\label{eq:relation3}
\end{equation}

Combining Equations (\ref{eq:kperp})--(\ref{eq:relation3}) and setting $v_\perp\sim v_z$\footnote{The linear dispersion relation yields $v_\perp\approx v_z(2\Omega/N)(k_z/k_\perp)$ for modes
with $\gamma\approx N_*$, and hence $v_\perp\sim v_z$ for wavenumbers given by Equations (\ref{eq:kz}) and (\ref{eq:kperp}).}
 yields the principal results:
the temperature gradient  
 is 
\begin{eqnarray}
\label{eq:ml1}
-{d\left<  T\right>\over dz}=N_*^2\sim {F^{2/5}\Omega^{4/5}\over H^{4/5}} \ ;
\end{eqnarray}
 the velocity and temperature fluctuations are
\begin{eqnarray}
v_z&\sim& {H^{1/5}F^{2/5}\over\Omega^{1/5}} \ ,
{\label{eq:ml2}}
 \\
T_1&\sim& {F^{3/5}\Omega^{1/5}\over H^{1/5}} \ ;
\label{eq:ml3}
\end{eqnarray}
and
the lengthscale of the modes that dominate heat transport is $1/k_\perp$ where
\begin{eqnarray}
k_\perp\sim {\Omega^{3/5}\over H^{3/5}F^{1/5}} \ .
\label{eq:ml4}
\end{eqnarray}

 Eqs. \ref{eq:ml1}--\ref{eq:ml4} are applicable 
for fast rotation ($\Omega\gg N_*$). 
For negligibly slow rotation,  one should
 replace $\Omega\rightarrow N_*$, yielding
   $N_*^2\sim F^{2/3}/H^{4/3}$, $v_z\sim (FH)^{1/3}$, and $k_\perp\sim 1/H$.
  These relations agree with 
the predictions of standard (non-rotating) mixing length theory. 
One may alternatively ``derive'' them by dimensional analysis.  However, 
they have proven more difficult to verify in simulations than their rotating counterparts. 
We defer further investigation of slowly rotating convection to future work.

The theory above was  proposed by \cite{Stevenson1979} 
(see his Eq. 43)
who additionally considered the case when rotation and gravity are misaligned,
 as well as magnetic fields.  
We have rederived it here in a slightly simpler way in order to highlight  
the main assumptions so that they may be tested in simulations. In addition, Stevenson's work
appears to be insufficiently
 well-known or  accepted in the literature.
 For example, \cite{Julien2012} derive Eq. \ref{eq:ml1} in a different way by scaling
 to the state at marginal stability---similar to the approach used by \cite{Kraichnan1962} for non-rotating convection. (Note that Eq. 2 in \cite{Julien2012} is equivalent to our Eq. \ref{eq:ml1} if one replaces $\Delta T/H\rightarrow N_*^2$ inside their dimensionless numbers, as is appropriate if most of the temperature drop is not within a boundary layer).
 And \cite{Aubert2001}  rederive Eq.~\ref{eq:ml2} (see their Eq. 10); note that \cite{Hide1974} Eq. A32 also gives the same prediction for the rms velocity.
  We suspect the main reason for this neglect is that Eqs. \ref{eq:ml1}--\ref{eq:ml4} have been 
insufficiently tested by numerical simulations---although the  simulations by \cite{Julien2012}
using a simplified set of equations provide  support, and
the prediction for the convective velocity has been compared with numerical simulations by \cite{Gillet2006} using a  quasi-geostrophic model. 
Note that
  \cite{2011Icar..211.1258S}
     suggest that a relation different than Eq. \ref{eq:ml2}
(specifically, $v_z\sim (F/\Omega)^{1/2}$) might be applicable.
The main goal of this paper is to verify Equations (\ref{eq:ml1})--(\ref{eq:ml4}) definitively with
high resolution simulations, which we turn to now.

\section{Simulations With Heating and Cooling Zones}
\label{Results}

As described in \S \ref{sec:intro}, RBC produces
thin boundary layers adjacent to the top and bottom plates through which heat enters and exits the fluid
via conduction.  As one decreases the diffusivities
$\nu$ and $\kappa$ towards  astrophysically tiny values, the boundary layers
become increasingly thin, and hence difficult to resolve in numerical simulations.  
But the  boundary layers themselves are of little interest if one's goal is to explain
convection zones in stars and planets.  
  To avoid this difficulty, we 
  directly heat a   zone of fluid at the bottom of the simulation domain
($z\in[0,0.2]$)
and cool a zone at the top ($z\in[1.2,1.4]$).
  That allows us to pass a flux of our choosing through the ``convection zone''
  ($z\in[0.2, 1.2]$) without suffering from overly thin boundary layers \footnote{ 
  Another possible route to avoiding boundary layers is to employ periodic boundary conditions  in the vertical
  direction  (e.g.~\citealt{Lohse2003,Garaud2010}).  However, such
  simulations develop large-scale up-flows that exit at
  the top of the box and enter at the bottom, as well as oppositely directed down-flows.
  These flows typically dominate the heat transport. 
    By contrast, in our setup  fluid is forced to turn around before reaching the top or bottom
  of the box, as in realistic convection zones.}.
  In the heating and cooling zones, the dynamics will depend 
  on the heating/cooling prescription.
  But  in the convection zone, the dynamics should be universal when 
  the diffusivities are small enough.  That will be tested both with convergence studies as
 $\nu$ and $\kappa$ are decreased, and with a few direct comparisons with
   standard Rayleigh-Benard simulations (in the next section).
  
  For most of our runs we integrate the equations of motion    (Eqs. \ref{eq:eoma1}--\ref{eq:eoma2} in the Appendix) with
   the pseudospectral code SNOOPY 
  \citep{Lesur2005}, and also rerun some of the simulations 
  with a
  very different code, the spectral element code  Nek5000 \citep{nek5000}, as an independent check.
  We set $\nu=\kappa$ throughout, and 
  boundary conditions are horizontally periodic and vertically impenetrable.
  Further numerical details are described in   \S \ref{methods}, including
  the implementation of heating/cooling, as well as a table of 
  simulations
  (Table~\ref{Table}). The choice $\nu=\kappa$ is made for simplicity. However, if the bulk properties of the convection are independent of the diffusivities (as predicted by Eqs.~\ref{eq:ml1}--\ref{eq:ml4} and verified
  below in our simulations), then they should not depend on this ratio.
   Nonetheless, it would be desirable to verify this directly with future simulations.

  We choose the amount of  heating/cooling so that the total heat flux
  (conductive + convective) through the 
   convection zone is unity in steady state.  The heat flux is defined as
   \begin{eqnarray}
   F&\equiv& {-\kappa}{d\left<T\right>\over dz}+\left< v_zT\right>,
   \end{eqnarray}
  where  angled brackets represent averaging in the horizontal direction at fixed height.  
  In steady state, the time-averaged $F$ is independent of height in the convection zone, 
  and we choose its value to be \textbf{ $F=1$}.
    That, together with our choice for the depth of the convection 
  zone ($H=1$), sets the simulation units\footnote{
  Heat flux
   has  units $[F]$= [{\it length}$]^2$/[{\it time}]$^3$ in the Boussinesq equations, and
   temperature has units $[T]=[{\it length}]/[{\it time}]^2$.}.
   Equivalently, lengths are measured in units of $H$ and time in units of $H^{2/3}/F^{1/3}$.
  The predictions of mixing length theory (Eqs. \ref{eq:ml1}--\ref{eq:ml4}) become particularly
  simple in our chosen units, and depend solely on $\Omega$.
 
\begin{figure}
  \begin{center}
   \hspace*{\fill}
    \subfigure{\includegraphics[trim=0.5cm 6cm 8cm 1cm, clip=true,width=0.48\textwidth]{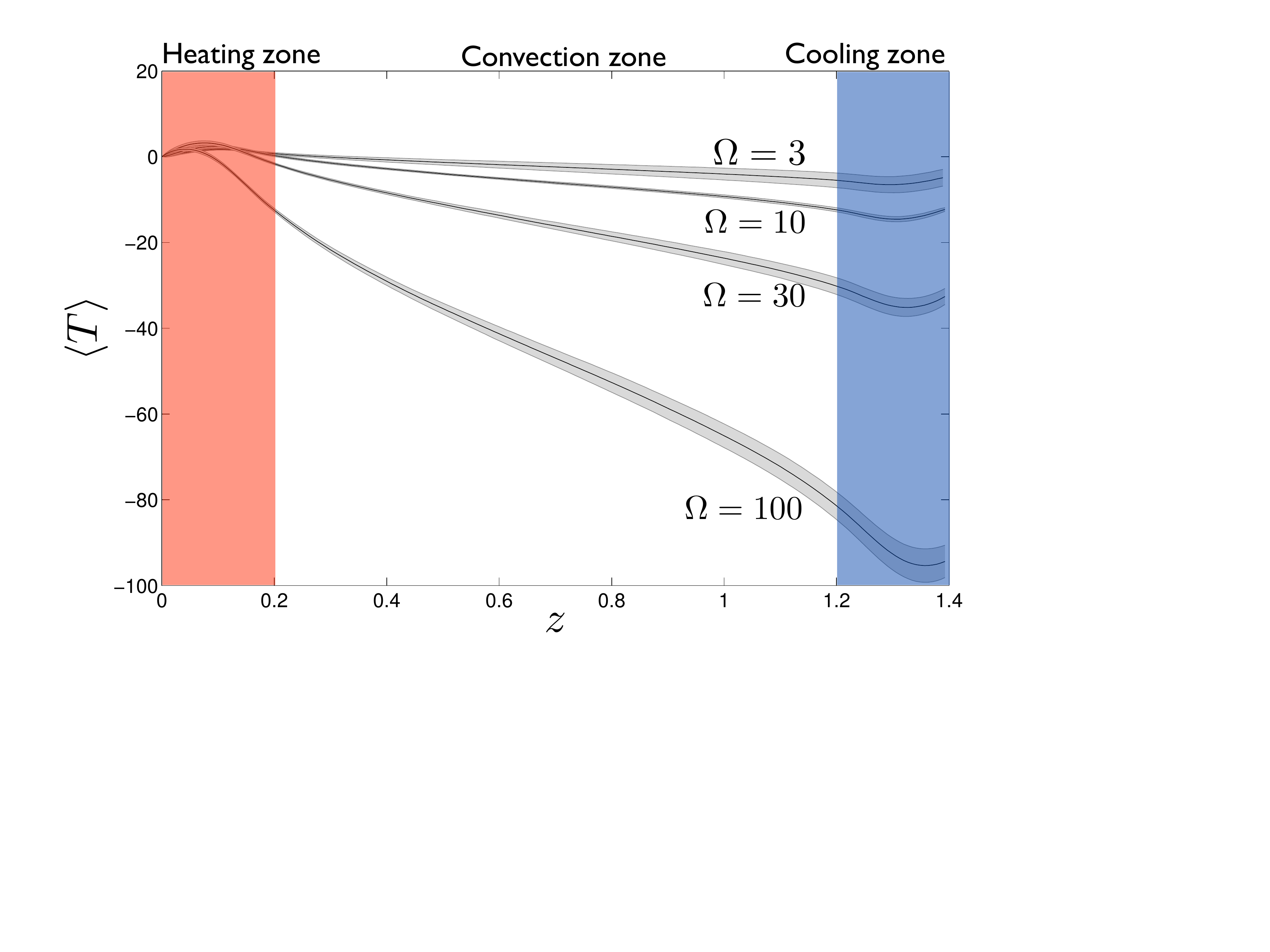}} \hfill
   \hspace*{\fill}
     \subfigure{\includegraphics[trim=0.5cm 6cm 8cm 1cm, clip=true,width=0.48\textwidth]{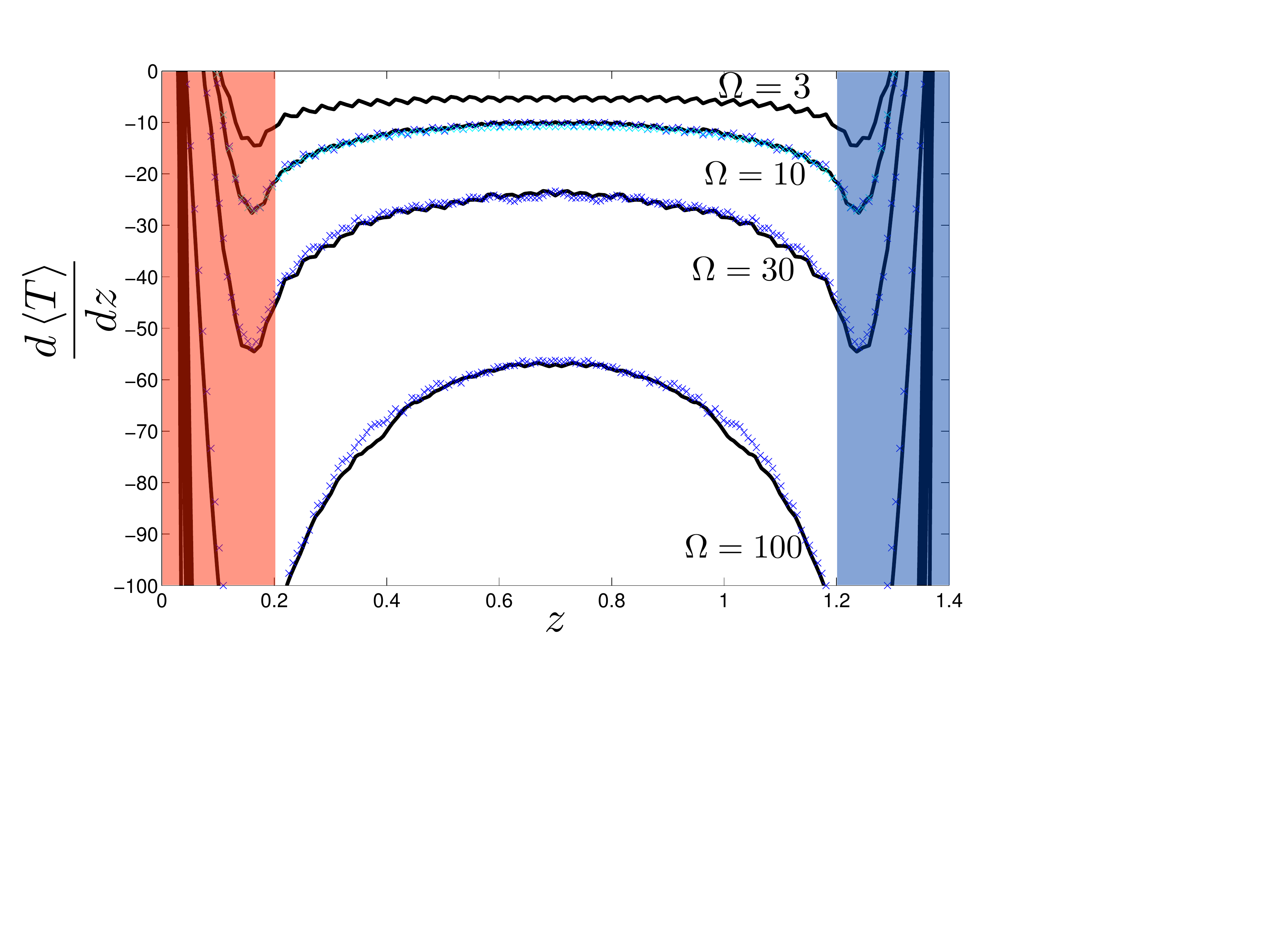}} \hfill
    \end{center}
  \caption{
   Top: 
  Steady-state temperature profiles in four simulations 
  (labelled 3M, 10M, 30L and 100L in Table~\ref{Table})
  with different values of $\Omega$, averaged over horizontal planes and in time. 
   The shading around each curve  represents the corresponding RMS fluctuation. 
Temporal averaging is over at least 50 time units.
   Bottom: 
   The black lines show the slopes of the four profiles in the top panel.
   Also shown as colored points are comparison simulations with 
   higher diffusivities (blue for ones labelled M in Table~\ref{Table} and light blue
   if labelled H).
  Our simulation units throughout this paper are determined by setting $F=H=1$.
  }
  \label{2}
\end{figure}

\begin{figure*}
  \begin{center}
    \hspace*{\fill}
        \subfigure{\includegraphics[trim=8cm 6.5cm 8cm 4cm, clip=true,width=0.46\textwidth]{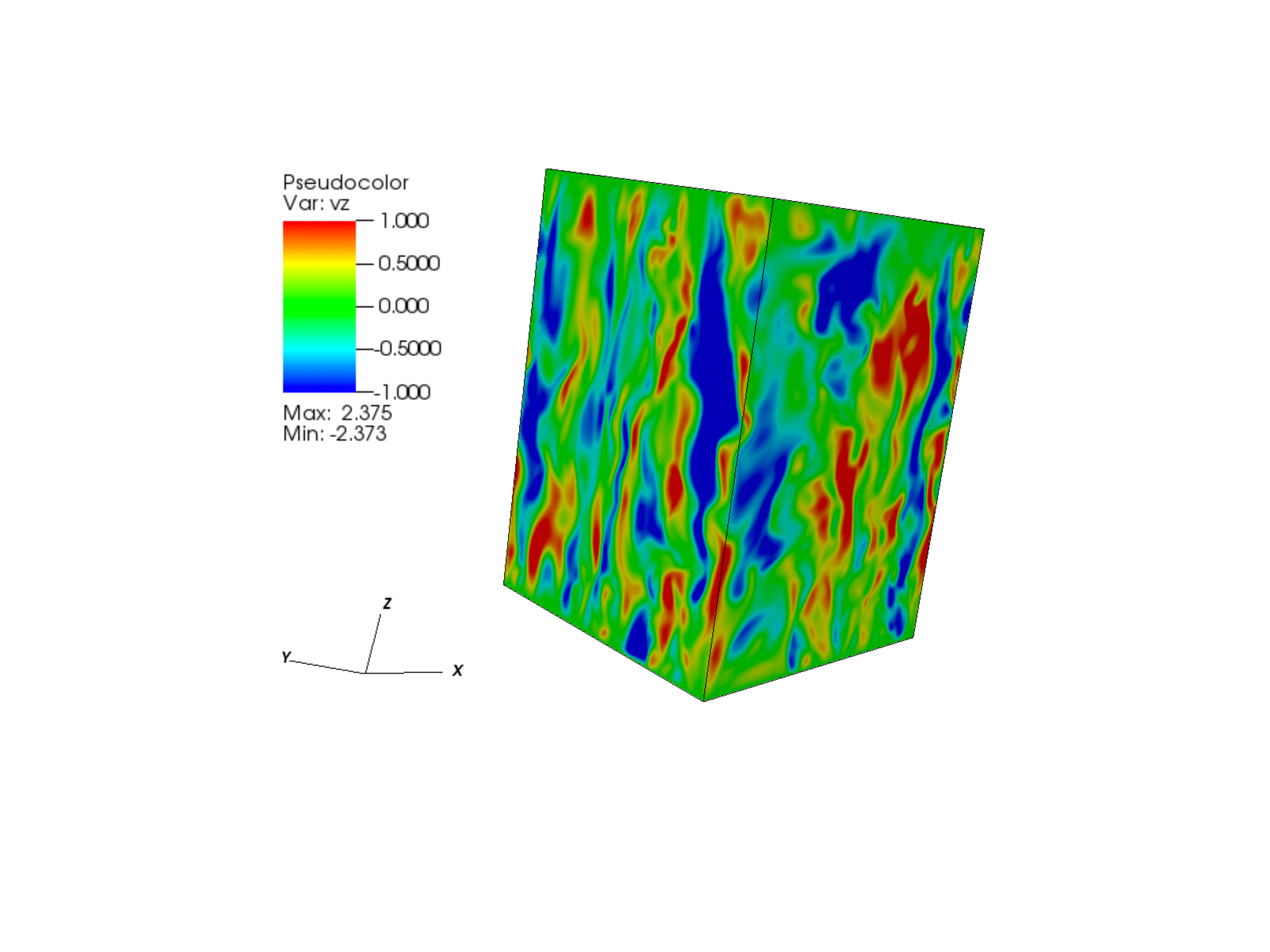} }
      \hspace*{\fill}
     \subfigure{\includegraphics[trim=8cm 6.5cm 8cm 5cm, clip=true,width=0.46\textwidth]{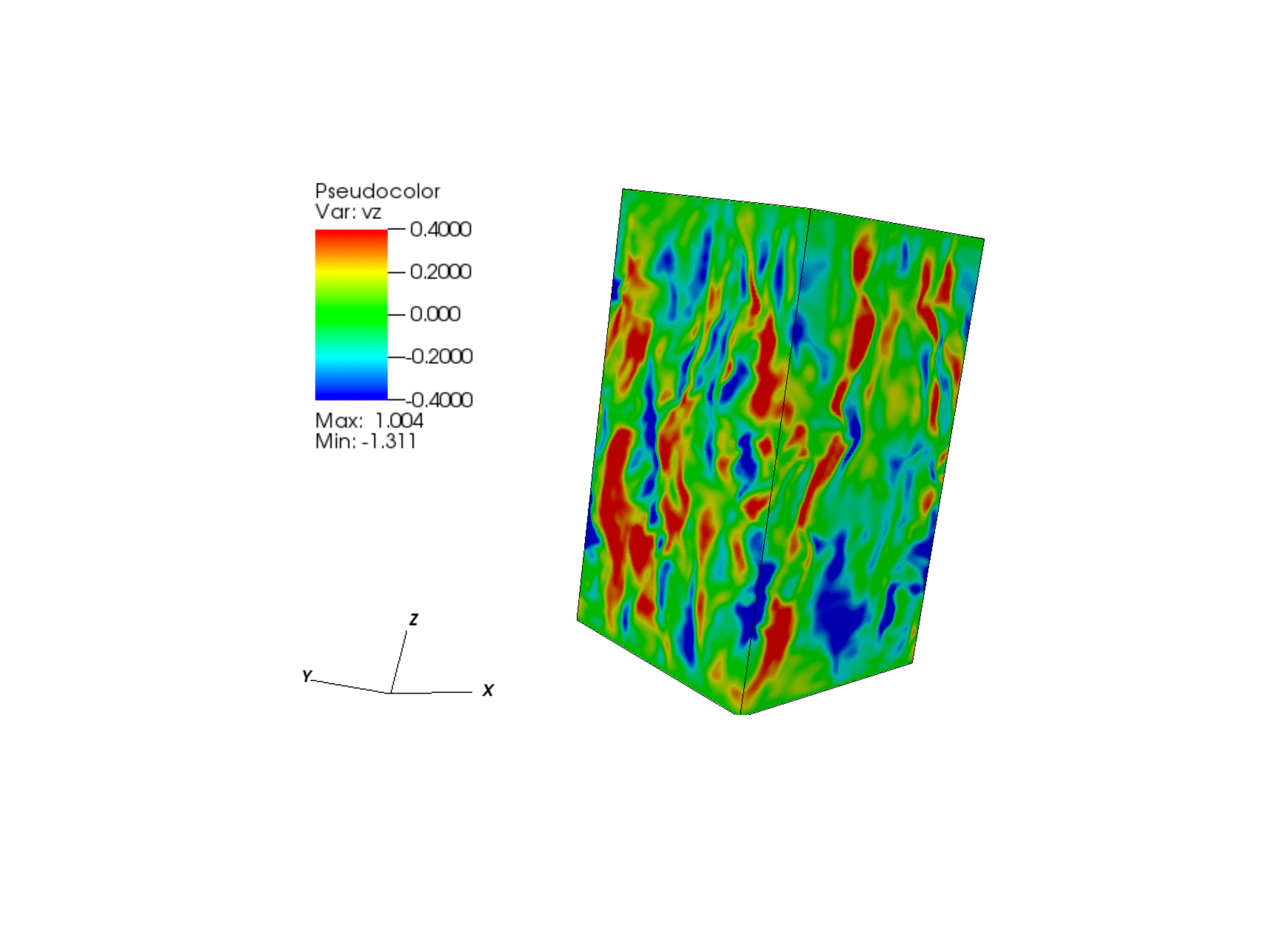} }
    \end{center}
  \caption{Vertical velocity in a simulation with $\Omega=10$ (left panel; simulation 10M) and
  $\Omega=1000$ (right panel; 1000L), both at $t=50$.
  In the right panel, the horizontal scale of the box has been stretched by a factor of $100^{3/5}$ relative to the left panel, and the colour scale has similarly been scaled by $100^{-1/5}$. These scalings correspond with the predictions in \S \ref{mlt}.}
  \label{1}
\end{figure*}

The top panel of Fig.~\ref{2} shows the  averaged 
temperature profiles
in steady state for four  simulations with different rotation rates.
 The temperature gradient in the convection zone
   steepens as $\Omega$ is increased. That is because rotation
   hinders convection: to achieve a given flux ($F=1$) requires a steeper gradient
   in a more rapidly rotating simulation. 
 The bottom panel of Fig.~\ref{2}  shows the slopes of the profiles in the top panel (black curves), 
 along with those from  simulations with higher diffusivities (colored points). 
The fact that the points agree with the  curves shows that
we are  probing the regime in which the bulk properties are independent (or at worst  weakly dependent) on the microscopic diffusion coefficients.\footnote{
The  importance of thermal diffusivity in the convection zone can be quantified by the ratio 
of conductive to total flux:
 $F_{\rm cond}/F=\kappa d\left<T\right>/dz$ (since $F=1$).  For example, in the 
 $\Omega=10$
 simulation, we find at the midplane $F_{\rm cond}/F \approx 10^{-3.3} \times 11\approx 0.006$. 
 The smallness of this ratio suggests that diffusivities play little role in the convection zone.
}

Fig.~\ref{1} shows a snapshot of $v_{z}$ in two simulations, 
one with $\Omega=10$ and the other with $\Omega=1000$.  
For the right panel, we  scaled the horizontal length scale, as well as the colour scale, by the amount predicted by the arguments in \S \ref{mlt}, relative to the left panel. 
 The similarity of the flow in both panels provides  support for the 
mixing length theory.
The dominant convective modes occur on smaller horizontal length-scales for more rapid rotation (Eq.~\ref{eq:ml4}), and the corresponding vertical velocities decrease (Eq.~\ref{eq:ml2}).

\begin{figure}
  \begin{center}
      \subfigure{\includegraphics[trim=0cm 0cm 0cm 0cm, clip=true,width=0.48\textwidth]{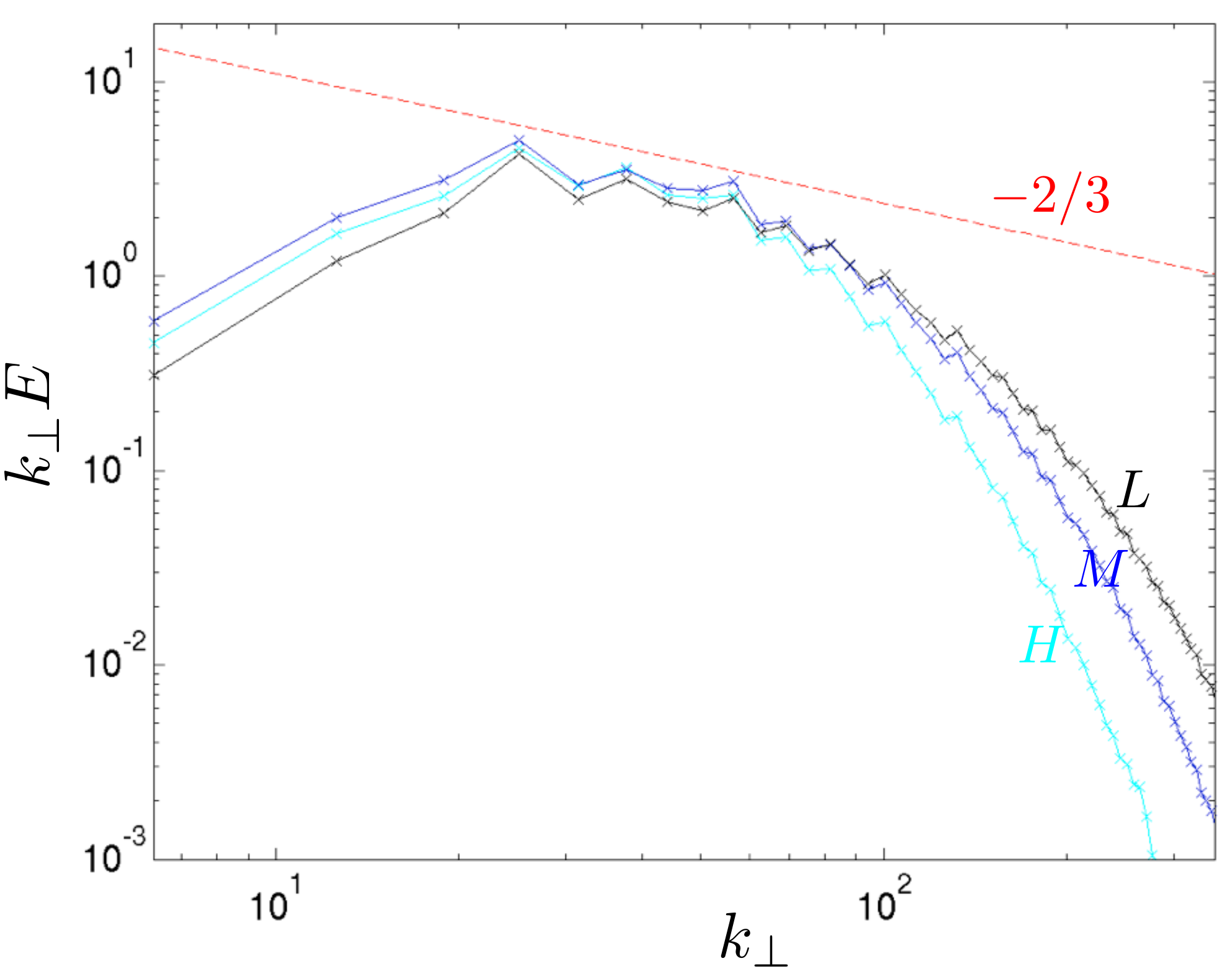} }
  \end{center}
  \caption{
   Spectrum of the heat flux for three simulations with 
   $\Omega=10$ and
   different viscosities.
   The ordinate is  $k_\perp E(k_\perp)= k_\perp^2
   \mathrm{Re}\left<
    \hat{v}_{z}(\bld{k}_{\perp},z,t)\hat{T}(\bld{k}_{\perp},z,t)^{*}\right>$ (within a constant), where hats denote Fourier transforms in the horizontal directions with horizontal wavevector $\bld{k_\perp}$, and
      the averaging is threefold: (i) over $z$  
      inside the convection zone;
      (ii) over the orientation of $\bld{k}_\perp$; and
       (iii)  over time until convergence.
       Note that the spatially averaged flux is $F={\rm const}\times\int E(k_\perp)dk_\perp$, 
       so that the peak of the spectrum plotted indicates the wavenumbers that dominate
       the flux. 
       The $-2/3$ line corresponds to the Kolmogorov scaling, i.e., 
        the assumption that 
         fluctuations across lengthscale $\lambda_\perp$ scale as
       $\delta v_z, \delta T\propto\lambda_\perp^{1/3}$, and that
       the two are strongly correlated. 
     The jaggedness is due to the discrete nature of the simulation grid, not temporal
     fluctuations which have been averaged away. }
 \label{3}
\end{figure}

 Fig.~\ref{3} shows the spectrum of the heat flux 
  in three $\Omega=10$ simulations that have different diffusivities. 
 The heat flux in these simulations is dominated by wavenumbers near $k_\perp\sim 20-30$. 
 When $\nu$ and $\kappa$ are decreased, the spectrum does not change near those lengthscales, indicating
 that the modes that dominate the heat flux are well-resolved
 and hence  bulk properties are independent of $\nu$ and $\kappa$.
The convection is anisotropic, since the depth of the convection zone is considerably
larger than the scale of modes that dominate the heat flux.  This anisotropy
is evident in Fig. \ref{1}, and is 
  expected from linear theory (Eq. \ref{eq:kperp}).
We also deduce from Fig.~\ref{3} that
 when $\nu$ is decreased, 
  the inertial range is extended to smaller scales.
   But those small scales  have little influence on the
larger scales that carry the bulk of the heat transport. 
These deductions conform  with the expectation from \S \ref{mlt}.
 
\begin{figure*}
 \begin{tabular}{l r}
 \subfigure{\includegraphics[trim=0cm 0cm 0cm 0cm, clip=true,width=0.48\textwidth]{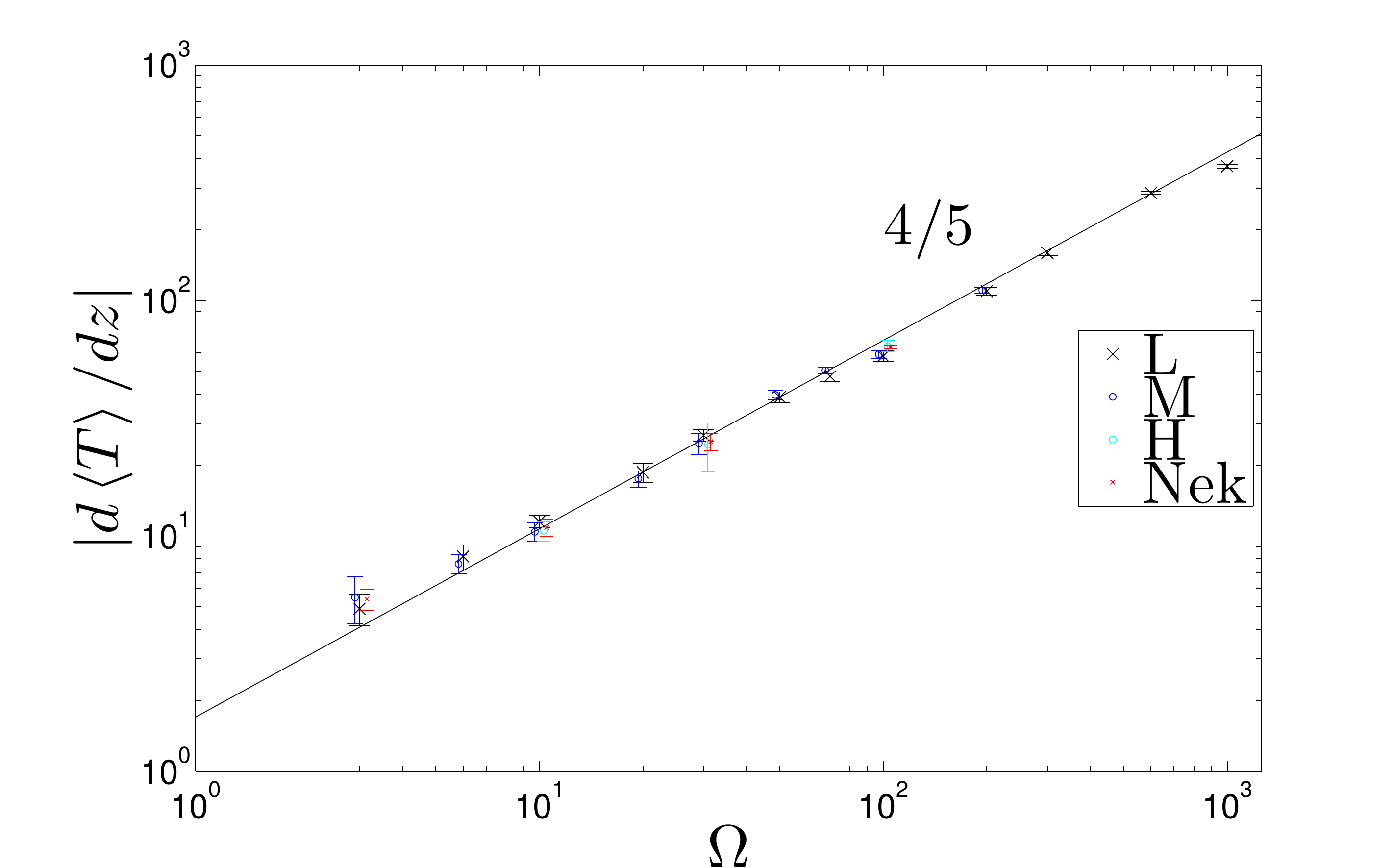} } &
  \subfigure{\includegraphics[trim=0cm 0cm 0cm 0cm, clip=true,width=0.48\textwidth]{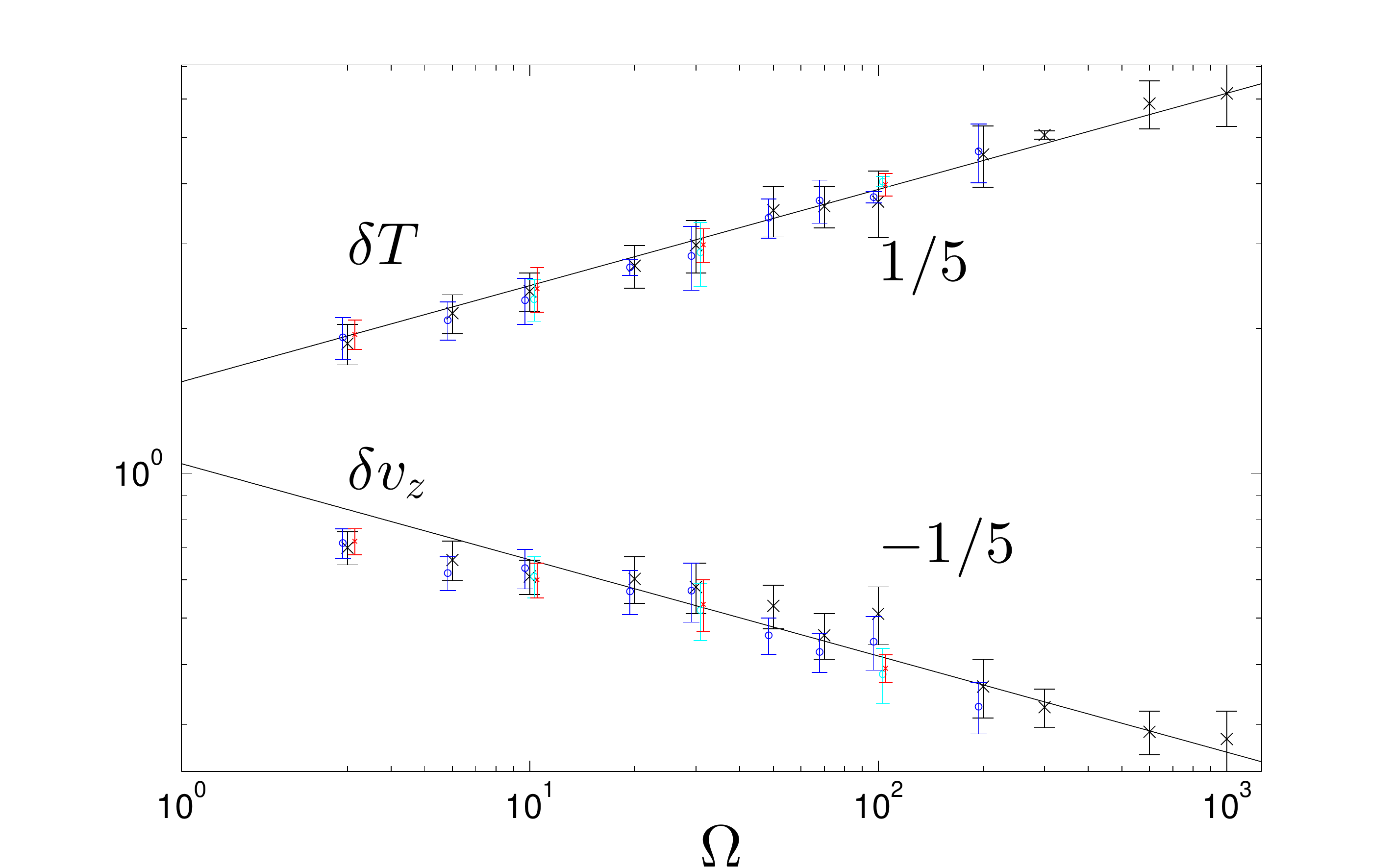} } \\
  \subfigure{\includegraphics[trim=2cm 0cm 0cm 0cm, clip=true,width=0.47\textwidth]{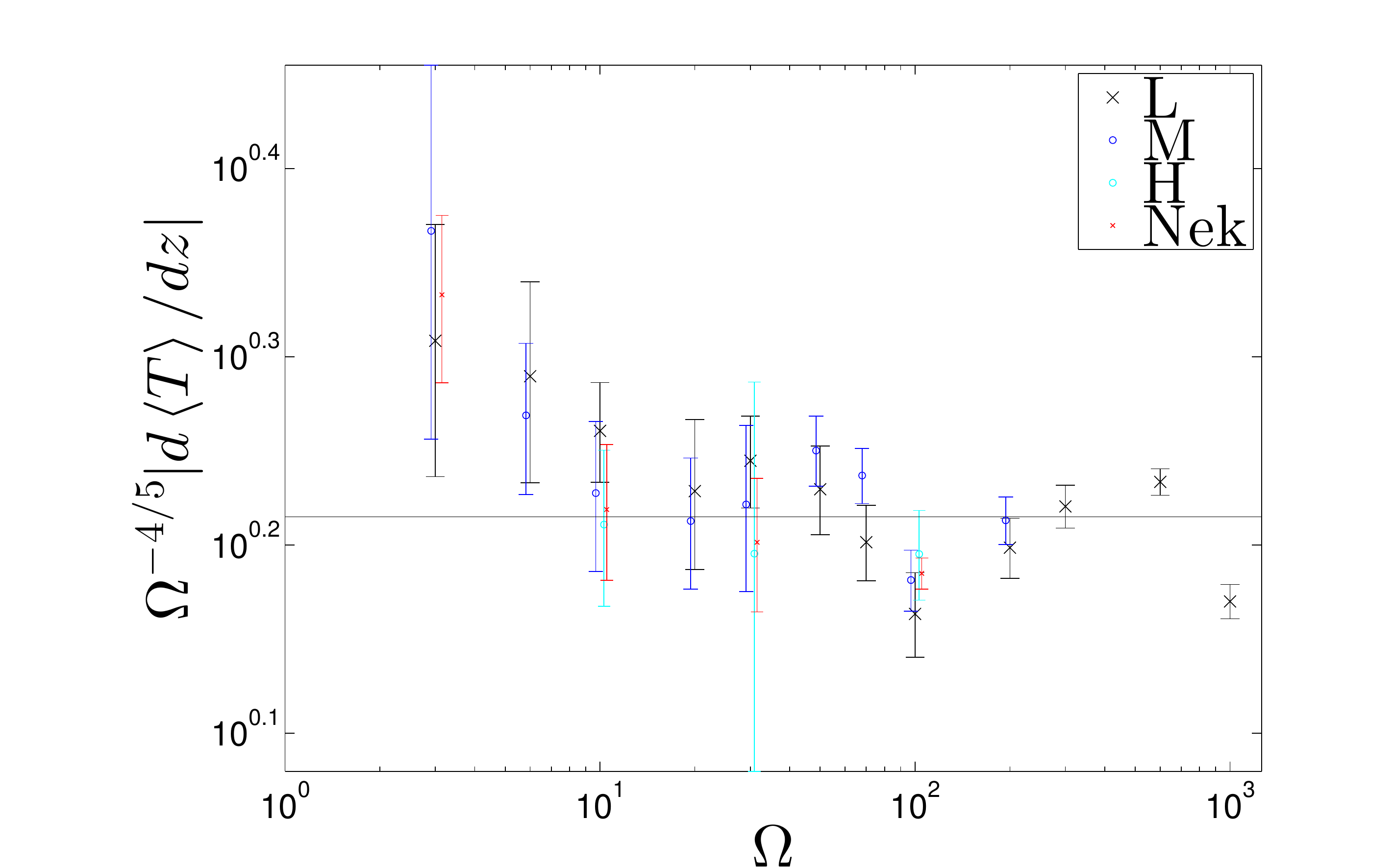} } &
   \subfigure{\includegraphics[trim=0cm 0cm 0cm 0cm, clip=true,width=0.48\textwidth]{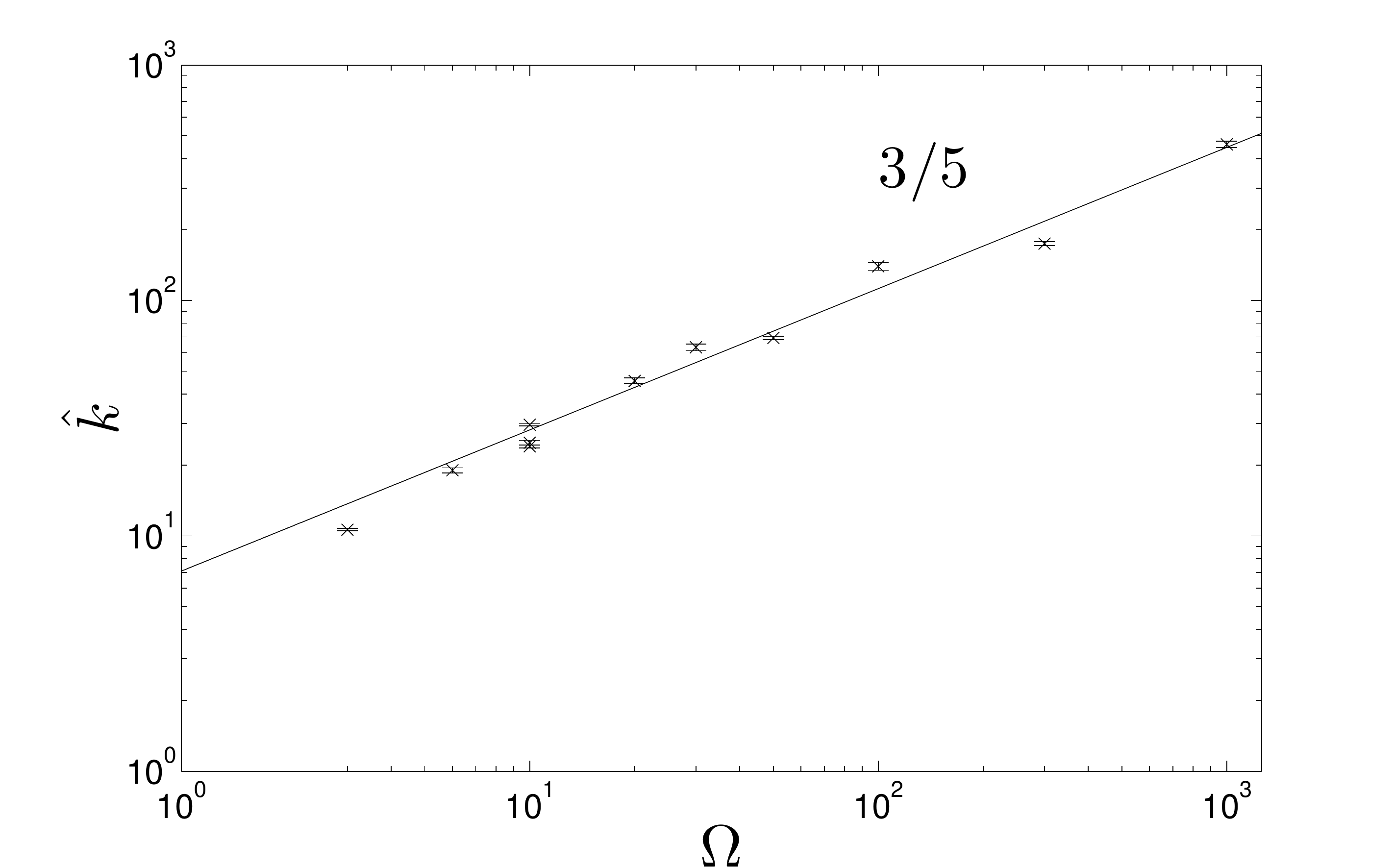} }
  \end{tabular}
  \caption{Dependence of  bulk properties   on  rotation rate.
   The predictions from mixing length theory  are shown as lines.
   The points labelled
    L, M, and H are SNOOPY simulations with viscosities from low to high; 
    red points are Nek5000 results; and error bars are RMS fluctuations. Points with the same $\Omega$ are slightly offset horizontally for clarity. Our simulation units are determined by setting $F=H=1$.}
  \label{5}
\end{figure*}

Figure~\ref{5}, the main result of this paper, displays the bulk properties 
in all of the simulations listed in Table \ref{Table}.
The top-left panel shows  the average temperature gradient in the convection zone, 
which 
 we extract  by fitting  the temporally and horizontally averaged temperature profile in the central $\frac{1}{3}$ of the simulation domain with a straight line. The gradient is shown as a point and the  corresponding RMS fluctuation as error bars, after averaging over at least 50 time units.
 At each $\Omega$, the results from the different simulations---SNOOPY with
 various viscosities and Nek---agree quite well  with the prediction, shown as a solid line.
 We fit the points and RMS errors 
 for all simulations
 with a linear least squares fit in log-log space, keeping points with
  $\Omega>6$ only (i.e.~the rapidly rotating limit).
 We find 
 \begin{equation}
  -{d\left<T\right>\over dz}=1.9^{+0.3}_{-0.3}\; \Omega^{0.75\pm0.06}.
  \end{equation}
   The exponent agrees with the theoretical prediction  of 0.8 (Eq. \ref{eq:ml1}) within 
   the error bars.  The lower-left panel of Fig.~\ref{5} plots the same data after removing the predicted scaling. 
 The prediction  works remarkably well
   over more than two orders of magnitude; exponents that differ from 0.8 by more than $\sim 10\%$ are definitively ruled out.

The top-right panel of Fig.~\ref{5} shows  the RMS fluctuations in $v_z$ and $T$ at the 
midplane of the box $(z=L_z/2)$. These also agree very well with the predicted
scalings, shown as lines.
A least-squares fit to the simulation points gives
\begin{eqnarray}
\delta v_{z}&=&0.9^{+0.5}_{-0.3} \; \Omega^{-0.15\pm0.08}\ ,
\\
\delta T &=& 1.5^{+0.9}_{-0.6} \; \Omega^{0.21\pm0.13} \ ,
\end{eqnarray}
the exponents of which may be compared with the predictions 
of -0.2 for $\delta v_z$ (Eq. \ref{eq:ml2}), and +0.2 for $\delta T$ (Eq. \ref{eq:ml3}).

In the fourth panel of Fig.~\ref{5}, we plot the horizontal wavenumber that dominates the heat flux, $\hat{k}$.  We define it via
\begin{equation}
\hat{k}={\int k_\perp E(k_\perp) d\ln k_\perp\over \int E(k_\perp)d\ln k_\perp} \ ,
\end{equation}
where the flux spectrum $E(k_\perp)$ is defined in the caption of Fig. \ref{3}.  Note that both the
numerator and denominator of this expression are dominated by large scales. 
A least-squares fit to the plotted values yields
\begin{equation}
\hat{k}=7.3^{+7.2}_{-3.6}\; \Omega^{0.58\pm0.19}  \ ,
\end{equation}
which may be compared with the mixing length prediction $k_\perp\propto \Omega^{0.6}$
(Eq.~\ref{eq:ml4}).

The results of this section, and in particular Fig.~\ref{5}, provides strong support for the
mixing length theory  presented in \S \ref{Theory}.

\section{Simulations Without Internal Heating and Cooling}
\label{RBC}

For the simulations presented in \S \ref{Results}, we directly heated and cooled the fluid 
inside the simulation domain to avoid thin boundary layers.
That approach was 
predicated on the assumption that the details of how
heat enters and leaves the convection zone is only of  minor importance for determining the bulk properties. In this section, 
we test that assumption.  To do so, we run comparison simulations 
without any internal heating or cooling.
The temperature
is held fixed at the bottom boundary and the flux is fixed at the top, which is essentially
the setup for standard Rayleigh-Benard convection. 
 Since the fluid is driven by heating at the boundaries, it is essential to correctly resolve the thermal boundary layers
  -- failing to do so results in an incorrect heat flux through the domain \citep{Groetzbach1983,Shishkina2010}. When the diffusivities are small, 
   the boundary layers become very thin, and hence  computationally costly to resolve.

\begin{figure}
  \begin{center}
 \begin{tabular}{l r}
 \subfigure{\includegraphics[trim=1cm 8cm 8cm 1cm, clip=true,width=0.48\textwidth]{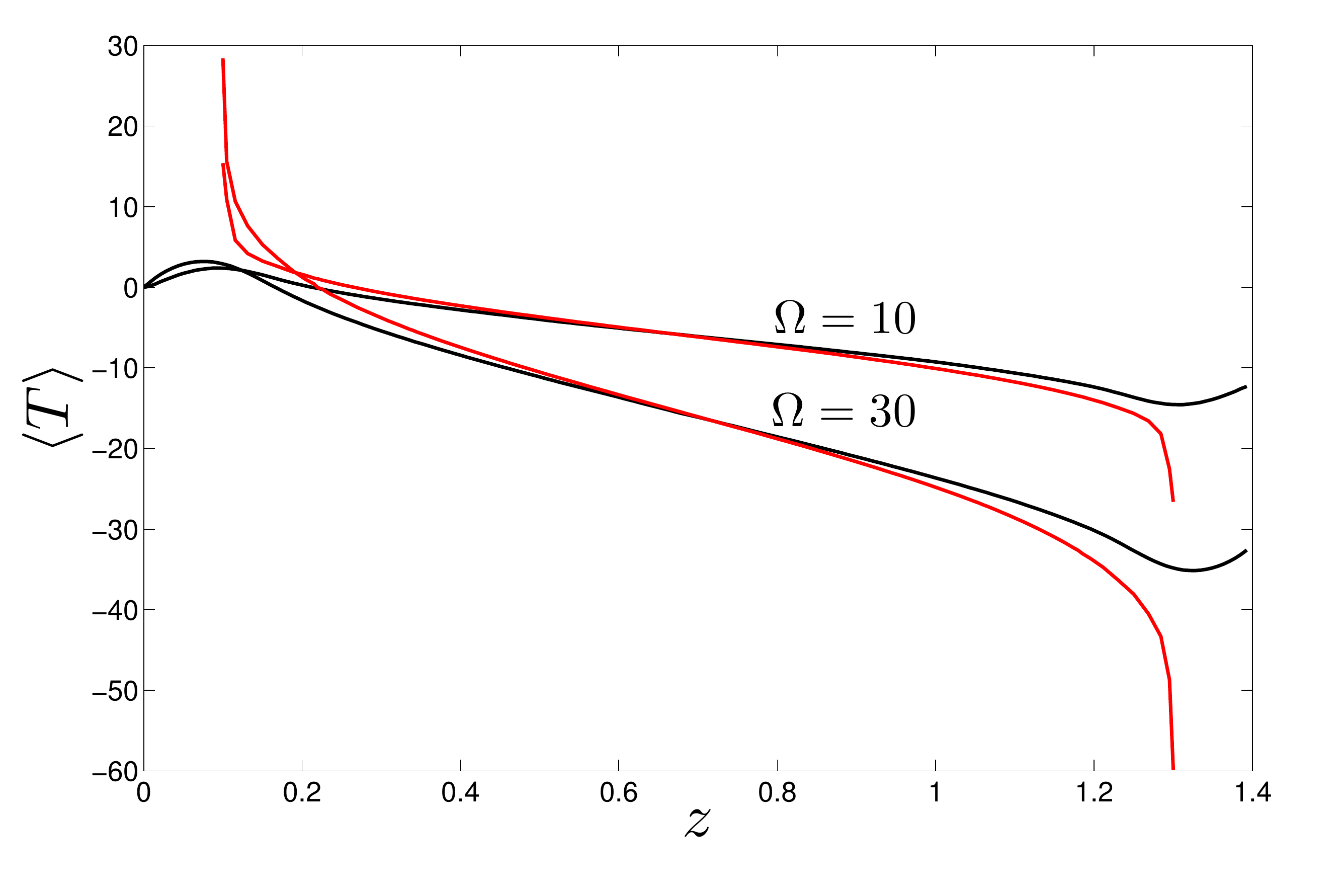} } \\
 \hspace{-0.4cm} \subfigure{\includegraphics[trim=0.6cm 8cm 8cm 2cm, clip=true,width=0.49\textwidth]{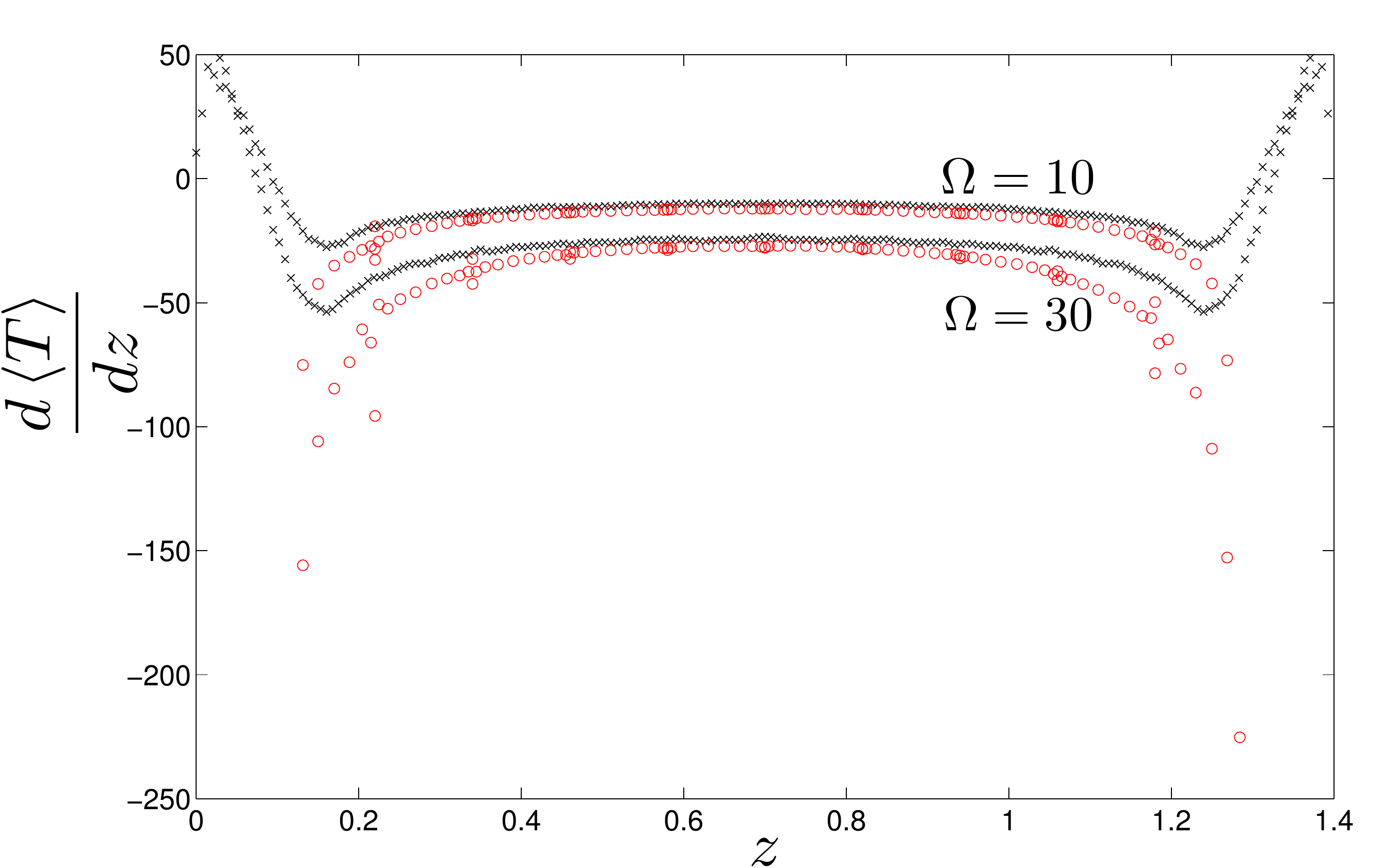} }
  \end{tabular}
     \end{center}
  \caption{
  Temperature profiles in simulations with (black) and without (red) internal heating at two 
  rotation rates.  Those with internal heating are 10M and 30M.  The ones with
  no internal heating were done with Nek5000. 
  The temperature was fixed 
  at the bottom and the flux fixed  at the top to $F=1$. We set $L_z=1.2$ in order to 
   crudely account for the fact that in the internally heated simulations the convection zone 
  extends somewhat into the heating/cooling zones. 
   The simulation without internal heating that has $\Omega=10$
   also has $L_{\perp}=1$,  $\nu=\kappa=10^{-3}$, and
  uses $200^3$ points split into 20 elements and 10th order polynomials within each element per dimension; the simulation with $\Omega=30$ has
  $L_{\perp}=0.5$, $\nu=\kappa=10^{-3.6}$, and similarly uses $200^3$ points.
 }
  \label{8}
\end{figure}

The equations of motion (Eqs. \ref{eq:eoma1}--\ref{eq:eoma2} with  $q=0$) were integrated
  with Nek5000, which is better suited than SNOOPY for resolving  thin boundary layers.   This is because grid points  in Nek5000 are clustered towards the boundaries, 
  whereas in SNOOPY sharp boundary layers produce unwanted Gibbs oscillations.
    Fig.~\ref{8} shows the temperature profiles and gradients from two such simulations,  with $\Omega=10$
and 30, and compares them with the corresponding heating/cooling zone simulations. 
The agreement
is  reasonable throughout the convection zone,  thus confirming our assumption\footnote{The agreement is not 
perfect primarily because of the difference in the depth of the convecting layer.}. The thin boundary layers are also evident in these figures.

\section{Discussion}
\label{Conclusions}

We  presented a simple derivation of mixing length theory 
in rapidly rotating convection, and then verified it with simulations.
The theory, postulated
by \cite{Stevenson1979},
 predicts the  properties of the convecting fluid under the assumption
that they are independent of microscopic diffusivities ($\nu$ and $\kappa$).
Equations \ref{eq:ml1}--\ref{eq:ml4} list the predictions for the mean temperature 
gradient,  the velocity and temperature fluctuations, and  the lengthscale of the modes that
dominate heat transport. 
Our simulation results, summarized in  
Figure \ref{5},  agree remarkably well with the theory, 
 across more than two orders of magnitude in rotation rate. 

We chose to focus on a very simple setup: Boussinesq convection in a box.
But despite its simplicity, and despite the vast literature already devoted to the topic,
 the result remains
under debate  \citep[e.g.,][]{King2012,Julien2012}, largely because of the complicating effect of
boundary layers. 
We circumvented this complication  by focusing 
on the properties of the convecting fluid---i.e., between the boundary 
layers.  
We did this by fixing the flux, 
and examining the  interior fluid's properties 
 for increasingly small  diffusivities.
 We thereby showed that the  convecting fluid's properties converged to the prediction 
of mixing length theory as $\nu$, $\kappa\rightarrow 0$.
Moreover, 
the numerical resolutions required to demonstrate convergence were relatively modest,
after artificially thickening the boundary layers with heating/cooling zones. 
  For example,
 our SNOOPY simulations had 256$^3$ gridpoints or fewer. 
Our numerical results  provide strong  support for  those   of  \cite{Julien2012}, who simulate
 a set of reduced equations  valid in the limit of rapid rotation.
 
Our work 
  lends confidence to mixing length theory's ability to accurately
 model highly turbulent convection. 
 We hope 
to extend it to include a variety of more complicated---and  realistic---effects, some
 of the most important of which are as follows:
\begin{itemize}
 	\item Including a background density gradient. One must then distinguish
	between entropy and  temperature.  The argument presented in \S \ref{mlt} should
	remain largely unchanged,  after replacing temperature with entropy.  
	But a possible complication is the asymmetry between upflows and downflows in the
	presence of 
	a density gradient (e.g.~\citealt{Hurlburt1984,Cattaneo1991,Miesch2005}).
	\item Allowing for a more realistic geometry, i.e., quasi-spherical rather than cubical.  
	The work presented in this paper strictly applies only to a small patch of the convective region near the poles of the star or planet. 
	An intermediate step before considering the full spherical problem would be
	 to allow rotation and gravity to be misaligned. \cite{Stevenson1979} predicts that in that case 
	Eqs. \ref{eq:ml1}--\ref{eq:ml4} should be altered by replacing $\Omega\rightarrow\Omega\cos\theta$, where $\theta$ is co-latitude. But that has yet to be confirmed by simulations.
	\item  Including the boundaries of a convection zone and the possibilities of 
	penetration and
	overshooting  into neighbouring stable layers (e.g.~\citealt{Hurlburt1994,Brummell2002,Rogers2005})
	\item  Allowing for the interaction between convection and differential rotation, and the generation of secondary flows.
	\item  Including  magnetic fields.
\end{itemize}

A large body of work has already been devoted to simulations of convection in rotating stars and planets, from Boussinesq (e.g.~\citealt{Hathaway1983,Tilgner2009,King2012}) to fully compressible \citep{Brummell1996,Brummell1998,Kapyla2005} Cartesian box simulations to Boussinesq \citep{Christensen2002,ChristensenAubert2006}, anelastic \citep{Glatz1984,Miesch2000,Kaspi2009,Jones2009,Gastine2012} and fully compressible simulations \citep{Kapyla2011} in spherical shell geometry.  Given the evident complexities of some of these simulations, it is our view that a more complete understanding of simpler models is required to enable us to understand these simulation results. Our work complements the literature by definitively verifying the rotating mixing length theory described in \S \ref{mlt} for the case of Boussinesq convection in the polar regions of a planet or star. We anticipate that the theory described in this paper, as well as the extensions discussed above, will help provide a theoretical basis for the simulation results.

Turning  to astrophysical applications of the theory, we note first that 
rotation
changes the entropy gradient relative to that
predicted by  standard (non-rotating) mixing length theory
by an order-unity factor---at least for the Sun, where
the rotation rate is comparable to the convective turnover time.
Thus the inclusion of rotation
 will not substantially change static
structure calculations, 
since 
it hardly affects the conclusion
that convection zones have a near-constant entropy throughout \citep{Stevenson1979}. But a potentially important application is explaining the differential rotation profile of the Sun and other stars. 
In particular, small latitudinal entropy gradients drive differential rotation via the thermal wind equation
 (e.g.~\citealt{Thompson2003,Miesch2006,Balbus2009a,Balbus2009b}). 
 Therefore to predict the differential rotation profile from first principles requires 
 one to understand how the entropy gradient depends on latitude.  It appears likely that 
 rotating mixing length theory (at least when extended to the case in which rotation and gravity are misaligned) will provide an important piece towards solving this puzzle.
 
 Another potential application is to tidal dissipation in a convective star or planet that has
  an orbiting companion.\footnote{We thank Jeremy Goodman for pointing out this application to us.}
This is important for understanding, for example, the tidal circularization of  solar-type binary stars out to approximately ten day orbits.
Previous  work estimates the turbulent viscosity due to convection by employing non-rotating
mixing length theory
\citep{Zahn1966,GN1977}.  
It would be of interest to see how the predictions are affected by employing rotating mixing
length theory.

\section*{Acknowledgments}
We  thank Jonathan Aurnou, Keith Julien and the referee for suggestions which have improved the manuscript.
YL acknowledges the support of NSF grant AST-1109776 and NASA grant NNX14AD21G. The computations in this paper were performed on Northwestern University's HPC cluster Quest.

\appendix

\section{Numerical Methods}
\label{methods}

\subsection{Convection with Heating/Cooling  Zones  (SNOOPY)}
\label{sec:ap}

The majority of our simulations  use
 SNOOPY (\citealt{Lesur2005,Lesur2010}), a
 Cartesian pseudo-spectral code.
  We use it to evolve the following equations of motion,
\begin{eqnarray}
\left(\partial_{t} + \boldsymbol{v}\cdot \nabla\right)\boldsymbol{v}  +2\Omega
\boldsymbol{e}_{z}
\times\boldsymbol{v}&=& -\nabla p +  T\boldsymbol{e}_{z} +\nu\nabla^2\bld{v},\ \ \ 
\label{eq:eoma1}
 \\
 \left(\partial_{t} + \boldsymbol{v}\cdot \nabla \right)T &=& q+\kappa\nabla^2 T,
\label{eq:eoma2}
\end{eqnarray}
which are modified from the strict Boussinesq equations (Eqs. \ref{eq1a}--\ref{eq1c}) by
 the inclusion of diffusive terms and
 a spatially variable heating/cooling function $q(z)$.
 Rather than evolving $T$ directly, we  write 
\begin{equation}
T=-\beta z+\theta,
\end{equation}
 with $\beta$ a constant, and evolve $\theta$.
 We describe below our choices
for $\beta$ and $q$. 
  Table \ref{Table} lists all of our heating/cooling simulations.
 
Throughout this paper we take $\nu=\kappa$, with a value as small as possible for a given number of grid points, subject to the constraint that the bulk properties be numerically well resolved.
 Our computational domain is a Cartesian box with dimensions $x,y\in [0,L_{\perp}]$ and $z\in[0,L_z]$, with $L_z=1.4$ throughout. 
  We vary $L_{\perp}$ until we have resolved the dominant convective scales, using intuition from linear theory \citep{C1961} and analysis of the horizontal energy spectrum. We verify that the bulk properties are independent of this parameter, once it is sufficiently large to capture the dominant convective modes. 
 
 Boundary conditions in the horizontal direction are periodic, and in the vertical direction are 
 impermeable ($v_{z}=0$), stress-free ($\partial_{z}v_{x}=\partial_{z}v_{y}=0$), and constant temperature ($\theta=0$) at the top and bottom ($z=0$ and $L_z$). Note that we evolve
 $\theta$ rather than $T$  because it allows us to impose 
the vertical boundary conditions on $\theta$ with a sine-wave decomposition.

\begin{table*}
\begin{center}
\caption{Simulation Parameters}
\begin{tabular}{ccccc||cccccc}
\tableline
\tableline
Label & $\Omega$ & $L_\perp$ & $-\log_{10}{\nu}$ & $N$ & $v_{z}$ & $|\frac{d\langle T\rangle }{dz}|$ & Ro$=v_{z}/(2\Omega H)$ & E$=\nu/(2 \Omega H^2)$ & $\mathrm{Ra}_f=FH^4/\nu^3$ & Nu$=F/(\nu |\frac{d\langle T\rangle}{dz}|)$ \\
\tableline
3M & 3 & 2 & 3 & 128 & 0.72$\pm 0.05$ & 5.5$\pm 1.2$ & $0.12$ & $1.7\times 10^{-4}$ & $1.0\times 10^{9}$ & 183 \\ 
3L & 3 & 2 & 2.7 & 192 & 0.70$\pm 0.06$ & 4.9$\pm 0.75$ & $0.12$ & $3.3\times 10^{-4}$ & $1.3 \times 10^{8}$ &102 \\ 
3Nek& 3 & 2 & 3 & 20$\times$10 & 0.72$\pm 0.05$ & 5.4$\pm 0.55$ & $0.12$ & $1.7\times 10^{-4}$ & $1.0\times 10^{9}$ &186 \\ 
6M & 6 & 1.5 & 2.6 & 128 & 0.62$\pm 0.05$ & 7.6$\pm 0.7$ & $5.2\times10^{-2}$ & $2.1\times 10^{-4}$ & $6.3\times 10^{7}$ & 53 \\ 
6L & 6 & 1.2 & 3.1 & 192 & 0.66$\pm 0.06$ & 8.2$\pm 1.0$ & $5.5\times10^{-2}$ & $6.6\times 10^{-5}$ & $2.0\times 10^{9}$ & 154 \\ 
10H & 10 & 1 & 3 & 128 & 0.61$\pm 0.06$ & 10.5$\pm 1.0$ & $3.0\times10^{-2}$ & $5.0\times 10^{-5}$ & $1.0\times 10^{9}$ & 95 \\ 
10M & 10 & 1 & 3.15 & 192 & 0.64$\pm 0.06$ & 10.4$\pm 0.95$ & $3.2\times10^{-2}$  & $3.5\times 10^{-5}$ & $2.8\times 10^{9}$ & 136 \\ 
10L & 10 & 1 & 3.3 & 256 & 0.61$\pm 0.05$ & 11.5$\pm 0.7$ & $3.0\times10^{-2}$ & $2.5\times 10^{-5}$ & $7.9 \times 10^{9}$ & 174 \\ 
10Nek & 10 & 1 & 3 & 20$\times$10 & 0.60$\pm 0.05$ & 10.9$\pm 0.9$ & $3.0\times10^{-2}$ & $5.0\times 10^{-5}$ & $1.0\times 10^{9}$ & 92 \\ 
20M & 20 & 0.7 & 3.3 &128 & 0.57$\pm 0.06$ & 17.5$\pm 1.4$ & $1.4\times10^{-2}$ & $1.3\times 10^{-5}$ & $7.9\times 10^{9}$ & 114 \\ 
20L & 20 & 0.7 & 3.55 & 192 & 0.60$\pm 0.06$ & 18.6$\pm 1.7$ & $1.5\times10^{-2}$ & $7.0\times 10^{-6}$ & $4.5\times 10^{10}$ & 191 \\ 
30H & 30 & 0.3 & 3.5 & 128 & 0.52$\pm 0.07$ & 24.4$\pm 5.7$ & $8.6\times10^{-3}$ & $5.3 \times 10^{-6}$ & $3.2\times 10^{10}$ & 130 \\ 
30M & 30 & 0.4 & 3.8 & 192 & 0.57$\pm 0.08$ & 24.7$\pm 2.5$ & $9.5\times10^{-3}$ & $2.6 \times 10^{-6}$ & $2.5 \times 10^{11}$ & 255 \\ 
30L & 30 & 0.4 & 4.05 & 256 & 0.58$\pm 0.07$ & 26.7$\pm 1.5$ & $9.7\times10^{-3}$ & $1.5\times 10^{-6}$ & $1.4\times 10^{12}$ & 420 \\ 
30Nek & 30 & 0.6 & 3.5 & 20$\times$10 & 0.53$\pm 0.06$ & 25.1$\pm 2.1$ & $8.9\times10^{-3}$ & $5.3\times 10^{-6}$ & $3.2 \times 10^{10}$ & 126 \\ 
50M & 50 & 0.5 & 3.5 & 128 & 0.46$\pm 0.04$ & 39.7$\pm 1.7$ & $4.6\times10^{-3}$ & $3.2\times 10^{-6}$ & $3.2\times 10^{10}$ & 80 \\ 
50L & 50 & 0.4 & 4.2 & 256 & 0.53$\pm 0.06$ & 38.8$\pm 2.1$ & $5.3\times10^{-3}$ & $6.3\times 10^{-7}$ & $4.0\times 10^{12}$ & 409 \\ 
70M & 70 & 0.4 & 3.5 & 128 & 0.43$\pm 0.04$ & 50.4$\pm 1.7$ & $3.0\times10^{-3}$ & $2.3 \times 10^{-6}$ & $3.2\times 10^{10}$ & 63 \\ 
70L & 70 & 0.35 & 3.75 & 192 & 0.46$\pm 0.05$ & 47.6$\pm 2.2$ & $3.3\times10^{-3}$ & $1.3\times 10^{-6}$ & $1.8\times 10^{11}$ & 118 \\ 
100H & 100 & 0.2 & 3.5 & 128 & 0.38$\pm 0.05$ & 63.9$\pm 3.5$ & $1.9\times10^{-3}$ & $1.6\times 10^{-6}$ & $3.2\times 10^{10}$ & 49 \\ 
100M & 100 & 0.2 & 4 & 192 & 0.45$\pm 0.05$ & 59.0$\pm 2.2$ & $2.2\times10^{-3}$ & $5.0\times 10^{-7}$ & $1.0\times 10^{12}$ & 170 \\ 
100L & 100 & 0.2 & 4.3 & 192 & 0.51$\pm 0.07$ & 58.0$\pm 3.0$ & $2.6\times10^{-3}$ & $2.5\times 10^{-7}$ & $7.9\times 10^{12}$ & 344 \\ 
100Nek & 100 & 0.2 & 3.5 & 20$\times$10 & 0.39$\pm 0.03$ & 63.4$\pm 1.2$ & $2.0\times10^{-3}$ &$1.6\times 10^{-6}$ & $3.2 \times 10^{10}$ & 50 \\ 
200M & 200 & 0.15 & 3.75 & 128 & 0.33$\pm 0.04$ & 111$\pm 3.2$ & $8.0\times10^{-4}$ & $4.4\times 10^{-7}$ & $1.8\times 10^{11}$ & 51 \\ 
200L & 200 & 0.15 & 4 & 128 & 0.36$\pm 0.05$ & 110$\pm 4.0$ & $9.0\times10^{-4}$& $2.5\times 10^{-7}$ & $1.0\times 10^{12}$ & 91 \\ 
300L & 300 & 0.2 & 4 & 128 & 0.33$\pm 0.03$ & 160$\pm 4.2$ & $5.4\times10^{-4}$ & $1.7\times 10^{-7}$ & $1.0\times 10^{12}$ & 63 \\ 
600L & 600 & 0.1 & 4.35 & 128 & 0.29$\pm 0.03$ & 286$\pm 4.6$ & $2.4\times10^{-4}$ & $3.7\times 10^{-8}$ & $1.1\times 10^{13}$ & 78 \\ 
1000L & 1000 & 0.05 & 4.7 & 128 & 0.28$\pm 0.04$ & 372$\pm 7.8$ & $1.4\times10^{-4}$ & $1.0\times 10^{-8}$ & $1.3\times 10^{14}$ & 135 \\ 
\tableline
\end{tabular}
\tablecomments{
Table of simulations with heating/cooling zones. The number under ``Label'' refers to $\Omega$, and the letters L, M, H, and Nek refer respectively to low, medium, and high viscosity SNOOPY simulations and to Nek5000 simulations. For SNOOPY simulations, the fifth column gives the number of Fourier modes in each direction, and for the Nek5000 simulations it gives the number of elements along each axis and the polynomial order within each element. Simulation parameters not listed in this table are given in \S \ref{sec:ap}. The data listed to the right of the vertical lines are derived from the simulation results. The sixth and seventh columns list the mean RMS vertical velocity and temperature gradient at $z=L_{z}/2$ with error bars, and the remaining columns list various non-dimensional parameters for comparison with other work. The bulk properties of these simulations are depicted in Fig.~\ref{5}. Our simulation units are determined by setting $F=H=1$. 
 To restore units one should replace $\Omega\rightarrow \Omega H^{2/3}/F^{1/3}$, $L_\perp\rightarrow L_\perp/H$, 
$\nu\rightarrow \nu/(F^{1/3}H^{4/3})$, $v_z\rightarrow v_z/(FH)^{1/3}$, and ${dT\over dz}\rightarrow {dT\over dz}H^{4/3}/F^{1/3}$. Note that $\nu=\kappa$ is assumed throughout.}
\end{center}
\label{Table}
\end{table*}

\begin{figure}
  \begin{center}
   \hspace*{\fill}
     \subfigure{\includegraphics[trim=0cm 0cm 0cm 0cm, clip=true,width=0.5\textwidth]{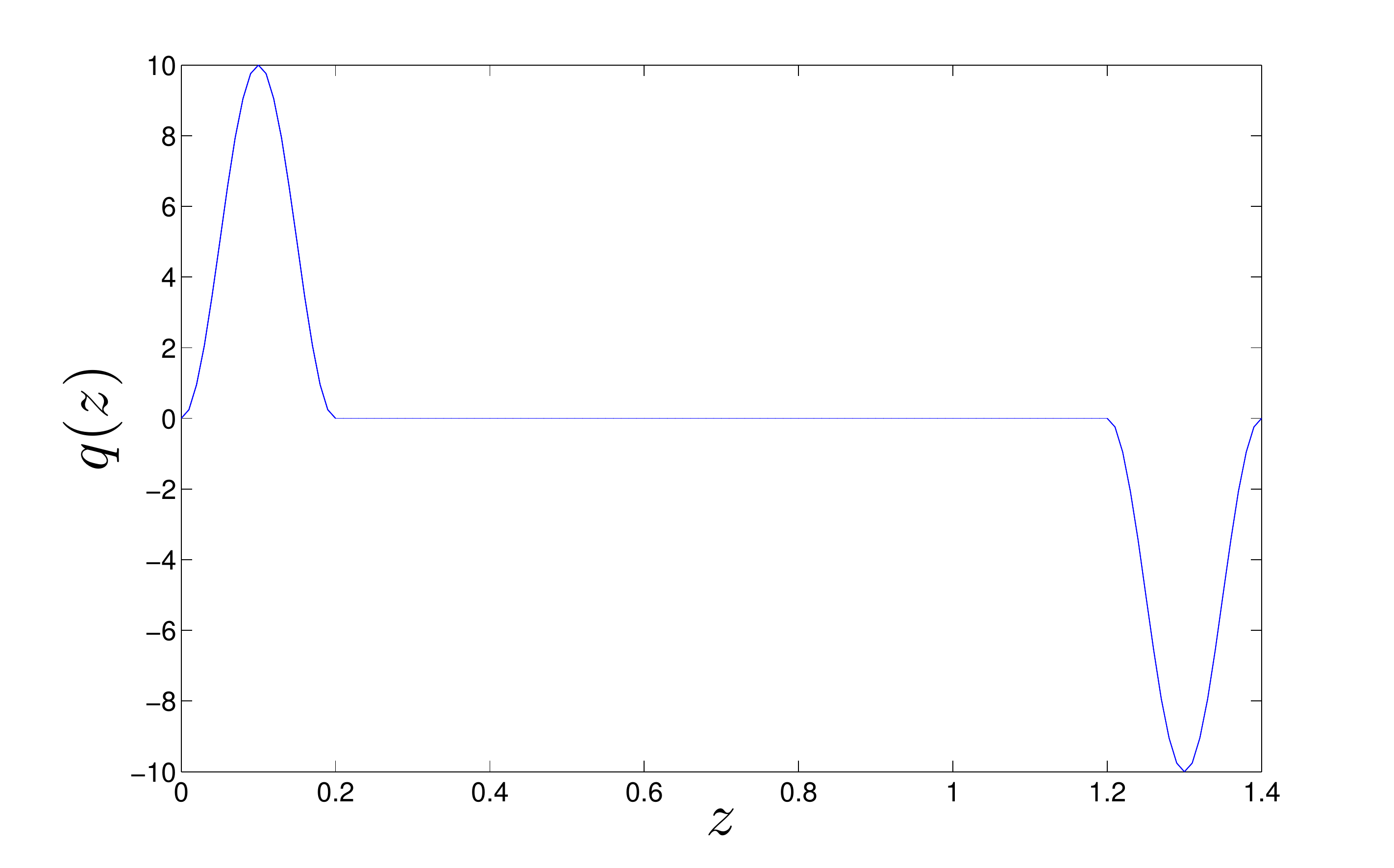} }\hfill
    \end{center}
  \caption{Internal heating and cooling profile used in our simulations. The convection zone  is the region $z\in[0.2,1.2]$, where $q(z)=0$.}
  \label{qheatcool}
\end{figure}

Our  $q$ (Fig.  \ref{qheatcool}) is chosen so that 
 fluid is heated at the bottom of the box
 in a zone of depth $\Delta=0.2$, and cooled by an equal amount at the top. 
 In addition, 
$q=0$
in
the central 
convection zone, which has depth $H=1$.
Explicitly,
\begin{eqnarray}
\label{hcprofile}
\nonumber
q(z) = \frac{1}{\Delta} \begin{cases}
 		  1+\cos \left(\frac{2\pi (z-\Delta /2)}{\Delta} \right) & \text{if } 0\leq z \leq \Delta , \\
   		 0  & \text{if } \Delta < z <L_{z}- \Delta , \\
		-1-\cos \left(\frac{2\pi (z-L_{z}+\Delta/2)}{\Delta } \right) &  \text{if } L_{z}-\Delta  \leq z \leq L_{z}, \\
  	 \end{cases}
\end{eqnarray}
which has integrated heating at the bottom
\begin{eqnarray}
\int_{0}^{\Delta}q(z) dz =1  \ ,
\end{eqnarray}
and an equal amount of cooling at the top. 
This gives unit flux
  in the convection zone  in steady state, as long as the flux vanishes at the top and bottom edges of the simulation box.

The value of $\beta$ is chosen to achieve zero flux at the box edges, using a relaxation method.
 Specifically, it
   is straightforward to show 
  that for the flux to vanish at the top and bottom edges
     one must have in steady state
     \begin{eqnarray}
  \kappa \beta L_{z}=L_z-\Delta-\int_0^{L_z}\left <v_z\theta \right>dz,
  \label{eq:k}
  \end{eqnarray}
  where the angled brackets here denote an average across a horizontal plane, as well as in time, and we have used the boundary
  conditions to eliminate some terms.    We evaluate the right-hand side of Eq. \ref{eq:k} in the code to give a target value for $\beta$, which we call $\tilde{\beta}$, and
relax $\beta$ towards its target value by solving
\begin{eqnarray}
\frac{d\beta }{dt} = \frac{1}{\tau}\left(\tilde{\beta}-\beta\right),
\end{eqnarray}
where $\tau$ is typically a few hundred in code units.
The simulation proceeds until Eq. \ref{eq:k} is satisfied within 1\%.
After that, $\beta$ is held fixed.
 In some of the simulations, turbulent fluctuations are sufficiently large that the relaxation method is never switched off according to the above criterion.
 This leads to slow and small changes in $\beta$.  By rerunning some of these simulations
 with the relaxation method turned off, we found that these slow, small changes do not
 appreciably change the mean properties (but they can amplify fluctuations by $\sim 50\%$).

We integrate the equations of motion with a first order splitting method, made up of
an implicit step for $q(z)$ and the diffusive terms, and  an explicit
 third order Runge-Kutta method for all other terms. 
The implicit step  uses
 an integrating factor to write
    the solution for a given Fourier mode at time $t_{n}$ as
\begin{eqnarray}
\hspace{-0.5cm}
\hat{\theta}(\boldsymbol{k},t_{n})=\hat{\theta}(\boldsymbol{k},t_{n-1})e^{-\kappa k^{2} \delta t}+\frac{\hat{q}(\boldsymbol{k})}{\kappa k^{2}}\left(1-e^{-\kappa k^{2} \delta t}\right),
\end{eqnarray}
where $\hat{q}$ is the discrete Fourier transform of $q$ and $\delta t$ is the time step. 
This  allows  larger timesteps to be used than with a fully explicit method.

\subsection{Nek5000}

We have also run a number simulations with Nek5000, 
an efficiently parallelised spectral element code \citep{nek5000}. 
For the simulations described in \S \ref{Results} with heating/cooling zones, 
the setup is almost the same as for the SNOOPY simulations described above, except we evolve 
$T$ directly rather than $\theta$. In addition, we
 impose zero flux ($\partial_{z}T=0$) conditions at each point on the top and bottom edges of the simulation box, rather than the constant temperature boundary condition in SNOOPY.  We also use explicit  second-order time integration for the heating and cooling terms,  whereas diffusive terms are integrated using an implicit method of the same order. The convective terms are fully de-aliased using the 3/2 rule, so that the polynomial order listed in Table.~\ref{Table} is actually $15$
 (not 10) for the integration of these terms.
  The simulations with Nek5000 agree well with those done with SNOOPY  (see Fig.~\ref{5}), which provides an independent check on our results.
  
\setlength{\bibsep}{0pt}
\bibliography{RB}
\bibliographystyle{apj}
\end{document}